\documentclass{aa}

\usepackage{graphicx}
\usepackage{txfonts}

\begin{document}



\title{Search for Double Transit Extrasolar Planetary Systems:
Another Transiting Planet Around OGLE-TR-111 or a False Positive Detection?}

\author{
Dante Minniti\inst{1}
}

\offprints{Dante Minniti}

\institute{
      Department of Astronomy, P. Universidad Cat\'olica, 
      Casilla 306, Santiago 22, Chile\\
      \email{dante@astro.puc.cl}
}

\date{Received .. ... 2005; accepted .. ... 2005}

\authorrunning{Dante Minniti}
\titlerunning{Search for Double Transit Extrasolar Planetary Systems}

\abstract{ 
The search for double transit planetary systems
opens new possibilities for the transit searches
and for studies of orbital stability, stellar irradiation, and migration 
scenarios, among others.
We explore the OGLE lightcurves of stars with confirmed planetary
companions (OGLE-TR-10, OGLE-TR-56, OGLE-TR-111, OGLE-TR-113, and OGLE-TR-132),
searching for additional transits.
The most promising candidate is OGLE-TR-111, where
the photometric measurements and the radial velocities 
are consistent with the presence of a second planet. 
If confirmed, OGLE-TR-111 would be the first extrasolar planetary system 
detected by transits.
The parameters of the possible new planet OGLE-TR-111c would be:
period $P = 16.0644 ~d$, semimajor axis $a = 0.12 ~AU$,
orbital inclination $i = 88-89 ~deg$, mass $M = 0.7 ~M_J$,
radius $R = 0.85 ~R_J$, density $\rho = 1.4 ~g/cm^3$. 
If confirmed, OGLE-TR-111c would be the smallest and densest extrasolar planet
measured todate, truly a Jovian planet,
with properties intermediate between Jupiter and Saturn,
albeit with shorter period.
Additional photometric and spectroscopic data would allow
to discriminate between a second transiting planet around 
OGLE-TR-111 and a false positive detection.
}

\maketitle

\section{Introduction: Extrasolar planetary systems}

The first "hot Jupiter" around the nearby solar-type star 51 Peg was discovered 
using precise radial velocity measurements (Mayor \& Queloz 1995).
The first multiple planet system around $\upsilon$ And was discovered also using
radial velocities (Butler et al. 1999). Indeed, this search technique
continues to be the most successful (Fischer et al. 2004, Pont et al. 2004,
Konacki et al. 2004).
The total sample of extrasolar planets todate discovered using the
radial velocity technique is reaching 135 planets (see the
Extrasolar Planets Encyclopaedia by J. Schneider 2005 
{\footnote{ http://www.obspm.fr/planets }}
).  In this sample there are 14
multiple planet systems, containing between 2 and 4 planets.  This means that
a significant fraction of the planets are in planetary systems.

A new search technique of transits has been systematically pursued by the OGLE
Collaboration, who published low amplitude ($\Delta I <0.08$ mag) 
transiting planetary candidates based on extensive $I$-band photometry
of several fields spread across the Milky Way
(Udalski et al. 2002a, 2002b, 2002c, 2003). While many of these
are binaries or blends, the OGLE sample contains real planets: five of
them have been confirmed using radial velocities 
(Bouchy et al. 2004, Pont et al. 2004, Konacki et al. 2003, 2004).  

Due to purely geometric effects, the discovery of 
short period transiting planets ($P<5$ days) is favoured.
One of the many advantages of the known transiting planets is that their
orbital plane is well defined. Using the reasonable assumption 
(unproved until now) that
other planets in these stars lie in the same orbital plane, one
can search for other transits. A major
 limitation is the quality of the photometry,
which demands confirmation of any other planets by radial velocities.

Here we report the search for double transiting planets in 
OGLE-TR-10, OGLE-TR-56, OGLE-TR-111, OGLE-TR-113, and OGLE-TR-132,
which are the OGLE systems confirmed todate to harbor single planets
(Table 1).
The initial search for additional unnoticed eclipses in the light curves
is followed by examination of the existing radial velocity data.
We find two possible interesting candidates, OGLE-TR-10 and OGLE-TR-111,
of which OGLE-TR-111 is the most promising one because its light
curve can be phased with a $P=16$ day period in absence of the known
planet OGLE-TR-111b. We conclude that further photometric coverage of
this system is needed.

\section{Search for Double Transiting Planets}

\begin{table}[t]\tabcolsep=1pt\small
\begin{center}
\caption{Sample OGLE Stars With Confirmed Planets.}
\label{Table1}
\begin{tabular}{l@{ }l@{ }l@{ }l@{ }l@{ }}
\hline
\multicolumn{1}{l}{OGLE}&
\multicolumn{1}{l}{RA}&
\multicolumn{1}{l}{DEC}&
\multicolumn{1}{l}{P(d)}&
\multicolumn{1}{c}{Reference}\\
\hline
&\\
TR10 & 17 51 28.25 & -29 52 34.9& 3.10139&Konacki+ 2005              \\
TR56 & 17 56 35.51 & -29 32 21.2& 1.21192&Konacki+ 2003              \\
TR111& 10 53 17.91 & -61 24 20.3& 4.01610&Pont+ 2004              \\
TR113& 10 52 24.40 & -61 26 48.5& 1.43248&Bouchy+ 2004, Konacki+ 2004              \\
TR132& 10 50 34.72 & -61 57 25.9& 1.68965&Bouchy+ 2004              \\
&\\
\hline
\end{tabular}
\end{center}
\end{table}

Systematics in the OGLE light curves are explored by Udalski et al. (2002a, 2002b, 2002c, 2003),
Drake (2003), and Sirko \& Paczy\'nski (2003). They discuss the possible
contaminants and different effects that are present in the OGLE transit sample.
We have decided to search for double transiting planets only on the 
stars that have been confirmed by radial velocities, where the parameters
of the planet are accurately known.
The stellar sample considered here is listed in Table 1, along with the main
properties of the transiting planets.
The most important parameter to consider when searching for additional planets 
is the orbital inclination, as the range of semimajor orbital axis of
additonal transiting planets depend strongly on this.
Even though the confirmed OGLE planets are very close to their parent stars,
fortunately, they have transits with flat bottoms, implying orbital inclination angles
$i>85$ deg, allowing the search for planets more distant from their stars.

The procedure followed is very simple, and can be described into four
main steps.

First, phasing the light curve of each star with the known period of the planet,
and substracting the points around the planetary transit. 
Typical OGLE light curves contain 1000-1500 data points, of which 70-140
around the transit are discarded. The light curves with the main transits
removed are called "reduced" light curves hereafter. We compute the dispersion
of these light curves by binning 10 contiguous datapoints, and check on the
stability of this dispersion (e.g. in other OGLE
light curves that turned out to be low amplitude grazing binaries
it was found that the dispersion increases).
This step was carried out for all the stars of Table 1.

Second, examination of this "reduced" light curve (Figure 1), to search for
asymmetries or excess points below the normal star baseline, beyond the 
photometric errors. 
This step was carried out for all the stars of Table 1.

Third, phasing the reduced light curve to search for periodicity that
may reveal an additional transiting planet. At this stage we pay particular
attention to periods which would be in mean motion resonances with the
original periods.
This step was carried out only for OGLE-TR-10 and OGLE-TR-111.

Finally, if a good candidate is found, we check the existing radial velocitiy
to confirm if they are consistent with the presence of an additional
planet, and try to derive its physical parameters from the photometry and spectroscopy. 
This step was carried out only for OGLE-TR-111

We discuss here the stars OGLE-TR-10, OGLE-TR-56, OGLE-TR-113, and OGLE-TR-132, reserving the next
section for a more thorough discussion of the most promising candidate, OGLE-TR-111.

%
\begin{figure}[h]
\resizebox{\hsize}{!}{\includegraphics{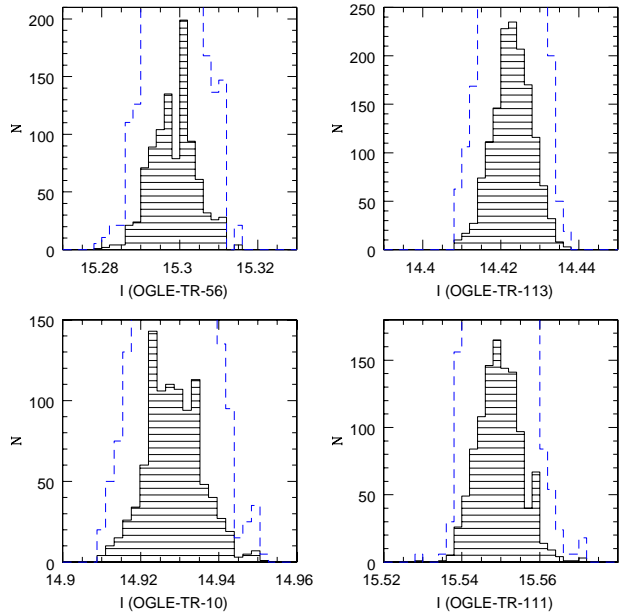}}
\caption{
Magnitude distribution for 
OGLE-TR-10, OGLE-TR-56, OGLE-TR-111, and OGLE-TR-113,
obtained after discarding the eclipses from the known transiting planets.
The open histograms are an expanded version of the hashed histograms
in order to illustrate possible excess of points below the mean 
magnitudes. Stars OGLE-TR-10 and OGLE-TR-111 show several points
4$\sigma$ fainter than the mean magnitudes.
}
\label{Fig01}
\end{figure}

\subsection{OGLE-TR-10}
OGLE-TR-10 shows larger than normal scatter compared with other OGLE stars at 
this magnitude ($I=14.93$). Its planet was suggested by
Bouchy et al. (2004), and was finally confirmed by Konacki et al. (2005). 
After substraction of the 117 points (from a total of 1082 points)
next to the transit of OGLE-TR-10b,
we find $\sigma_I=0.006$ mag. There are a group of fainter points suggesting
the possibility of additional transits (Figure 1), including
3 points located $4\sigma$ below the mean magnitude.
However, the reduced light curve could not be phased adequately to reveal any
clear low amplitude transits. More accurate photometric follow up of this star is needed.

\subsection{OGLE-TR-56}
OGLE-TR-56b was the first secure 
OGLE planet: it was confirmed as a planetary mass object by the
radial velocities of Konacki et al. (2003).
We substracted 137 points (out of 1115 total points) next to the transit of OGLE-TR-56b.
The scatter in the reduced light curve of OGLE-TR-56 gives $\sigma_I=0.005$,
smaller than OGLE-TR-10 in spite of being $0.4$ mag fainter.
However, the magnitude distribution of the reduced light curve is very symmetric,
and there is no evidence for points from additional transits.

\subsection{OGLE-TR-113}
OGLE-TR-113b was confirmed as a planet almost simultaneously by two different groups:
Bouchy et al. (2004) and Konacki et al. (2004).
The reduced light curve was obtained by eliminating 67 points centered on the
OGLE-TR-113b transits (out of a total of 1517 points).
The scatter in the reduced light curve of OGLE-TR-113 is very low, $\sigma_I=0.004$.
However, we find evidence for a long term periodic behaviour with $P=26.5$ d, at
very low amplitude ($0.005$ mag). If this is taken into account, the light curve
becomes even tighter. This is a good case of a star where strong limits may be put
on the absence of additional transiting giant planets with orbital semimajor
axis $a<0.2$ AU.

\subsection{OGLE-TR-132}
OGLE-TR-132b was confirmed with the radial velocities of Bouchy et al. (2004).
OGLE-TR-132 is the faintest star of this sample. 
We substracted 71 points (out of 1044 total points) next to the transit of OGLE-TR-132.
The dispersion of its light curve is large,  $\sigma_I=0.008$,
preventing us from putting useful limits to the presence of additional transiting planets.

\section{The Case of OGLE-TR-111}
\subsection{Stellar parameters for OGLE-TR-111}

We have previously carried out a selection of the most promising
OGLE planetary candidates using low dispersion spectroscopy in combination
with optical and near-infrared
photometry (Gallardo et al. 2004).  This work identified
OGLE-TR-111 as one of the most likely candidates to host exoplanets. The
planet OGLE-TR-111b was discovered by Pont et al. (2004)
using precise velocities.

Based on the spectroscopy, Pont et al. (2004) derive the following
stellar parameters for OGLE-TR-111:
mass $M = 0.82 M\odot$,
radius $R = 0.85 R\odot$,
temperature $T_{eff} = 5070$ K, 
gravity $log ~g = 4.8$, and
metallicity $[Fe/H]=0.12$ dex.

Based on a low resolution spectrum plus optical and infrared photometry,
Gallardo et al. (2004) derive the following stellar parameters:
radius $R = 0.71 R\odot$,
temperature $T_{eff} = 4460$ K,
reddening $E(B-V) = 0.16$,
absolute magnitude $M_V = 6.82$, and
distance $D = 850$ pc for OGLE-TR-111.

Even though the values from these independent studies are not identical,
both studies agree in the parameters within the uncertainties.
In order to derive planetary parametes hereafter 
we use for consistency the stellar parameters adopted by Pont et al. (2004),
with the cautionary remark that the major uncertainties on the planetary
parameters (mostly the radius) would arise from the uncertainties in the stellar properties.

\subsection{Photometric Evidence for OGLE-TR-111c}

The Existing planet OGLE-TR-111b is a massive planet with mass $M_p sin~i=0.53 ~M_J$,
radius $R = 1.0 R_J$, in an orbit with
period $P=4.0166$ days and semimajor orbital axis $a=0.047 ~AU$
(Pont et al. 2004).  They call this planet OGLE-TR-111b the "missing link".
The transit signature of this planet is clearly seen in the
phased OGLE lightcurve (Figure 2).

Aside from the
transit of OGLE-TR-111b, additional points well below the
baseline are observed, clumped at a single phase. Specifically,
there are 7 points located $4\sigma$ below the mean magnitude. This clumping hinted 
at the presence of additional transits with 
a period multiple of $P=4.011$ dat, the OGLE-TR-111b period.

Because the additional feature is 180 degrees out of phase we have to
make sure that it is not a noisy secondary eclipse caused by an
eclipsing binary system.
This possibility is discarded because of (a) the new transits are not always
present, and (b) the radial velocities of Pont et al. (2004) show
low amplitudes.

%
\begin{figure}[h]
\resizebox{\hsize}{!}{\includegraphics{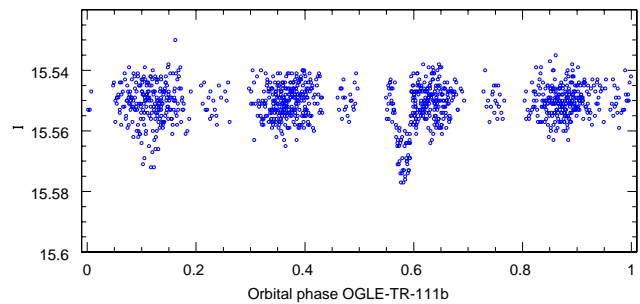}}
\caption{
OGLE lightcurve for OGLE-TR-111 phased for $P=4.0166$ days
(1176 photometric points), showing the transit of OGLE-TR-111b
at phase 0.6. Several points clumped at phase 0.1 hint at the presence
of additional eclipses.}
\label{Fig01}
\end{figure}

There are nine transits of OGLE-TR-111b observed by OGLE.
Looking at the unphased light curve (Figure 3), it became clear
that there might be some possible additional transits interleaved with 
some of the transits of OGLE-TR-111b. Two of these transits are
indicated with the arrows in Figure 3.

Figure 3 also clearly illustrates the fortunate fact that the period of
OGLE-TR-111b is a multiple of the Earth's rotation period, as discussed
by Pont et al. (2004), resulting in
a favorable transit observing condition during the OGLE observations. 

We then proceeded to phase the OGLE data in search for additional transits.
For this, 68 points around the OGLE-TR-111b
transits were discarded, phasing only the remaining 1108 baseline points.
A new phased possible transit was found with a period of $P=16.0644$ days, 
as shown in Figure 4. 
This period would in 1:4 mean motion resonance with
the previously known planet of this system OGLE-TR-111b, with $P = 4.0161 ~d$.
Caution should be taken with this period, because it is largely based on
two transit like features. The aspect of these two individual
transits is not different from the other nine transits of OGLE-TR-111b,
but a larger number of individual transits must be observed in order
to confirm this periodic nature.

Pending confirmation of this possible period, we note that a 
1:4 mean motion resonance should not be totally unexpected.
Let us consider
a simple example from our own neighborhood: the Jovian system is a good
example, because it is considered to be a stable "mini Solar system".
The Galilean satellites
Io and Ganimede share a 1:4 mean motion resonance.
But also Europa has a 1:2 mean motion
resonance with Io (and 2:1 with Ganimede), so on a very speculative
side one could search for
another planet around OGLE-TR-111 in a similar mean
motion resonance in between the two possible planets, with a period of $P=8.03$
days.  In the case of the  OGLE-TR-111 with an orbital inclination
of 88-89 deg, planets located as far out as 0.3 AU might be detected. This would
correspond to the 3:7 resonance
with the outer planet (e.g. Callisto is in 3:7 mean motion resonance with Ganimede).
With the more accurate periods to be determined by this years OGLE campaign,
one can tune the searches for transits from
additional planets to specific mean motion resonances. 

The possible transit appears to be symmetric, with a duration of
 $t_T=4$ hours. The transit also appears to
have a flat portion lasting about $2$ hours, ruling out a grazing eclipse,
and constraining the orbital inclination angle to $88-89$ degrees.
Other transient features such as star spots cannot be ruled out with the
available data. However, it is very suggestive that the two 
additional transits alternate with the main transits of OGLE-TR-111b
(Figure 4). 

%
\begin{figure}[h]
\resizebox{\hsize}{!}{\includegraphics{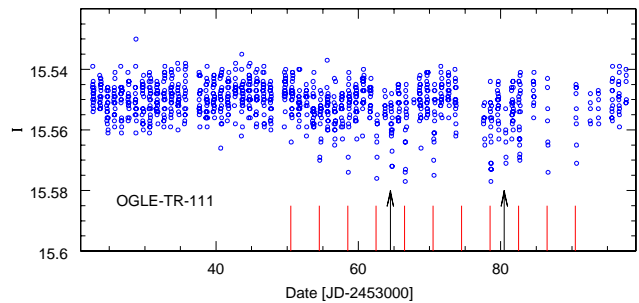}}
\caption{
Unphased OGLE data for OGLE-TR-111, showing the individual
transits of OGLE-TR-111b indicated by the vertical lines.
Two additional eclipses can be identified at the locations of the arrows.
}
\label{Fig02}
\end{figure}

%
\begin{figure}[h]
\resizebox{\hsize}{!}{\includegraphics{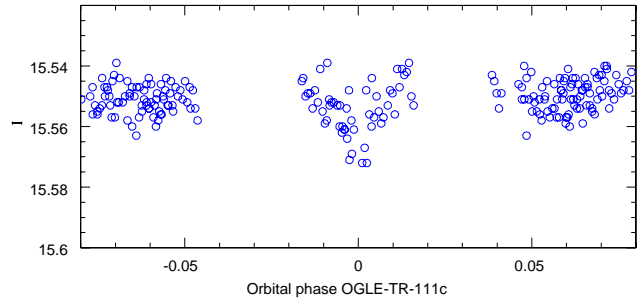}}
\caption{
Lightcurve of OGLE-TR-111 phased with $P=16.0644$ days
after discarding 68 points around the transits of OGLE-TR-111b.
The possible 4 hour long transit of OGLE-TR-111c is seen.
}
\label{Fig03}
\end{figure}

\subsection{Spectroscopic Evidence for OGLE-TR-111c}

The photometry of OGLE-TR-111 suggests the presence of an additional 
giant planet with $P=16.0664$ days, but it is not conclusive. 
Here we check if the existing
spectroscopic data are consistent with this suggestion, and  try to
derive the parameters of the possible additional planet.
We would call this possible extrasolar transiting planet OGLE-TR-111c, {\it a.k.a.}
the "missing link's sister".

Based on the radial velocities of the star, it is straightforward to fit the
eight individual velocity measurements of Pont et al. (2004)
with a double Keplerian model instead of a single one
(Figure 4). Given the original parameters for OGLE-TR-111b, 
this fit is not free, however: the period and the phase of
OGLE-TR-111c are fixed by 
the photometric data. The amplitude is a free parameter, which is clearly
poorly constrained due to the limited orbital coverage:
only one half of the putative second planet orbit is covered.
With this caveat, if there is an additional planet,
we obtain the parameters listed in Table 2. This Table presents the
photometric results and the final parameters for the
possible OGLE-TR-111c planet as well.
Different fits changing the original parameters for OGLE-TR-111b
are allowed, but are not significantly better.

The amplitude of the radial velocities induced by the additional planet are
scaled by the factor $M_P/M_*$. In the case of
OGLE-TR-111c, the fit shown in Figure 4 has 
velocity semiamplitude $V_r=60$ m/s, and $M_P=0.75 ~M_J$.
For comparison, the O-C residuals of the single planet fit are
$24$ m/s, which are reduced to $14$ m/s with the double Keplerian model.

The major difficulties for better constraining the parameters of the
two possible planets
circling the star OGLE-TR-111 are the uncertainty in the photometry and
the limited spectroscopic coverage.
Clearly, further photometric and spectroscopic observations over
an extended period of time are needed
to confirm the existence of OGLE-TR-111c, and to refine the measured parameters
of the planets.

Finally, the planetary radius listed in Table 2 is also uncertain
because it depends on the adopted stellar size.
In particular, adoption of a different stellar radius, $R = 0.71 R\odot$
from Gallardo et al. (2004), leads to a smaller planet radius $R = 0.7 ~R_J$ for
OGLE-TR-111c, which would translate into a higher density.

%
\begin{figure}[h]
\resizebox{\hsize}{!}{\includegraphics{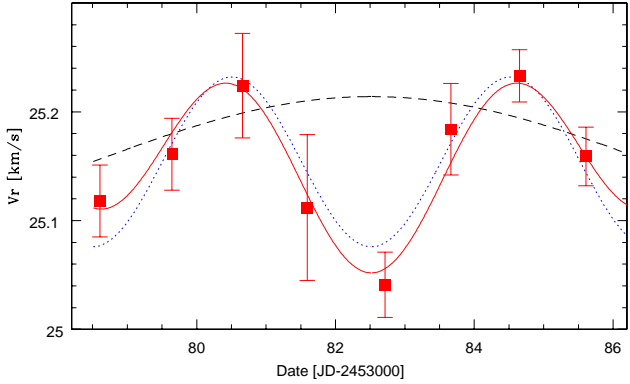}}
\caption{
Radial velocity measurements for OGLE-TR-111 from Pont et al. (2004).
The dotted line shows their fit for OGLE-TR-111b, the dashed line shows the
effect of OGLE-TR-111c, and the solid curve the combined effect of the
two planets.
}
\label{Fig01}
\end{figure}

\begin{table}[t]\tabcolsep=1pt\small
\begin{center}
\caption{Possible OGLE-TR-111c planetary parameters.}
\label{Table2}
\begin{tabular}{l@{ }l@{ }}
\hline
\hline
&\\
Orbital period       & $16.0644 \pm 0.0050 ~d$             \\
Semimajor axis       & $0.12 \pm 0.01 ~AU $              \\
Orbital eccentricity & $ 0 $ (assumed)          \\
Transit epoch        & $JD~2453064.73\pm 0.01 $          \\
Transit duration     & $4 \pm 1 ~h$                   \\
Transit depth        & $0.01 \pm 0.005 ~mag$              \\
Orbital inclination  & $88-89 ~deg$              \\
Systemic velocity    & $25.40  \pm 0.05~km/s$            \\
RV semiamplitude     & $60 \pm 20 ~km/s$               \\
Planet mass          & $M = 0.7 \pm 0.2 ~M_J$           \\
Planet radius        & $R = 0.85 \pm 0.15 ~R_J$          \\
Planet density       & $\rho = 1.4  \pm 0.3 ~g/cm^3$     \\
&\\
\hline
\end{tabular}
\end{center}
\end{table}

\subsection{Discussion of an Optimistic Scenario}

We have argued that the evidence is not conclusive to distinguish
between a second transiting planet around OGLE-TR-111 and a false positive detection.
However, here we briefly discuss the implications of a positive confirmation
of the existence of OGLE-TR-111c. If confirmed, this new planet would be unique and hold
several interesting records:

$\bullet$ OGLE-TR-111 would be the first extrasolar system with multiple
planets detected by transits, and the first multiple system for which
planetary masses and radii are measured.
With a distance of 850 pc, OGLE-TR-111 it would also be the most distant
extrasolar planetary system discovered todate.

$\bullet$ The transit of OGLE-TR-111c allows to constrain further the
inclination of the orbital plane of  the system to $i = ~88-89 deg$.
This would prove for the first time that the orbits of two planets of
an extrasolar planetary system are coplanar to within 2 degrees.
The most important implication of coplanarity is that careful photometric
monitoring may reveal rocky planets in this and other systems
by timing the transits of giant planets Holman \& Murray (2004), or by directly
detecting their transits.

$\bullet$ With radius $R = 0.85 ~R_J$, mass $M = 0.7 ~M_J$,
and density $\rho = 1.4 ~g/cm^3$, the planet OGLE-TR-111c would be
smallest and densest extrasolar planet found. It could be called a true
Jovian planet, albeit with shorter period,
because it has size like Saturn, density like Jupiter, and mass
intermediate between them. Its radius, however, could be as small as $R = 0.7 ~R_J$.

$\bullet$ With $P = ~16.0644$ d,
OGLE-TR-111c would also the transiting planet with the longest orbital period,
lying at the arbitrary boundary between "hot" and
"normal" giant extrasolar planets. 
In that sense it could also be considered a
"missing link" on its own, justifying the "missing link's sister"
nickname given here.

$\bullet$ Comparison with the properties of OGLE-TR-111b would directly
confirm the effects of inflation due to stellar
irradiation, as the less massive planet closer to the star would be bigger than
the more massive OGLE-TR-111c.
In fact, planet OGLE-TR-111b would be 1.22 times bigger than OGLE-TR-111c.
This result is independent of the uncertainties in the stellar mass and radius.
However, reducing the uncertainties in the radii is very important in
order to test the models.

$\bullet$ If the period of $P = ~16.0644$ d
is confirmed,
the OGLE-TR-111 system would an interesting case for stability studies because of the
presence of a 1:4 mean motion resonance. 
The confirmation of the orbital resonances coupled with the low inclination
of the system the exiting possibility of the search
for transiting smaller (rocky?) planets in OGLE-TR-111.

\section{Conclusions}

We have started a search for second planets around existing
transiting systems. This search has been applied
to the specific case of the stars 
OGLE-TR-10, OGLE-TR-56, OGLE-TR-111, OGLE-TR-113, and OGLE-TR-132.
Even though  OGLE-TR-56,  OGLE-TR-113, and OGLE-TR-132 do not show
evidence for scatter above what is expected, there are
several data points of OGLE-TR-10 and OGLE-TR-111 below their mean
magnitudes that suggest the possibility of additional transits.

While it was not possible to phase the OGLE-TR-10 reduced light curve 
in order to find the orbital period for another planet, the
OGLE-TR-111 reduced light curve was successfully phased with $P\approx 16$ days. 


Thus we explore the possibility of a putative additional planet in this system.
Based on published photometry and radial velocities, we 
tentatively derive the following parameters for OGLE-TR-111c:
orbital period
$P = ~16.0644$ d,
semimajor axis $a = 0.12 ~AU$,
mass $M = 0.75 ~M_J$,
radius $R = 0.85 ~R_J$, and
density $\rho = 1.5 ~g/cm^3$.

We stress that the possibility of a false positive detection is not ruled out,
and that this must be confirmed with additional data.
The major difficulty for securing the present claim of another planet
transiting around OGLE-TR-111 is the uncertainty in the photometry and
the limited spectroscopic coverage. Both difficulties would
be easily overcome with more observations.  We thus stress the need 
for the photometric and spectroscopic follow up of this system.

However, we speculate that extensive transit searches by space missions (KEPLER,
COROT, MPS) will find numerous multiple systems, as predicted by Holman \& Murray (2004).
The present search for double transit planetary systems suggests a number of
interesting follow-up studies. 
Theoretical modeling of formation and stability
of such systems including different migration scenarios can be pursued.
These systems would provide also a good opportunity to test various
models of the effects of stellar radiation in planetary atmospheres.
Further observation of multiple transit systems would refine the 
planetary parameters and reveal similarities and differences between
the structures of "hot" and "normal" extrasolar giant planets.

\begin{acknowledgements}
DM is supported by Fondap Center for Astrophysics 15010003.
\end{acknowledgements}

\end{document}